\RequirePackage{lineno}
\documentclass[12pt,superscriptaddress]{revtex4-1}
\usepackage[english]{babel}
\usepackage{indentfirst} 
\usepackage{graphics} 
\usepackage{epstopdf}
\usepackage{epsfig}
\usepackage{float}
\usepackage{chngcntr}

\newfloat{sup}{thp}{lop}[section]
\floatname{sup}{Supplementary Material}

\begin{document}
\title{Surface curvature guides early construction activity in mound-building termites}
\date{}
\author{Daniel S. Calovi}
\thanks{These two authors contributed equally}
\thanks{Corresponding authors: daniel.calovi@gmail.com, paulmb@ufl.edu}
\affiliation{Harvard School of Engineering and Applied Sciences, 33 Oxford Street, Cambridge MA 02138}
\affiliation{Wyss Institute for Biologically Inspired Engineering, 60 Oxford Street, Cambridge MA 02138}
\author{Paul Bardunias}
\thanks{These two authors contributed equally}
\thanks{Corresponding authors: daniel.calovi@gmail.com, paulmb@ufl.edu}
\affiliation{Department of Environmental and Forest Biology, SUNY College of Environmental Science and Forestry Syracuse, New York 13210}
\author{Nicole Carey}
\affiliation{Harvard School of Engineering and Applied Sciences, 33 Oxford Street, Cambridge MA 02138}
\affiliation{Wyss Institute for Biologically Inspired Engineering, 60 Oxford Street, Cambridge MA 02138}
\author{J. Scott Turner}
\affiliation{Department of Environmental and Forest Biology, SUNY College of Environmental Science and Forestry Syracuse, New York 13210}
\author{Radhika Nagpal}
\affiliation{Harvard School of Engineering and Applied Sciences, 33 Oxford Street, Cambridge MA 02138}
\author{Justin Werfel}
\affiliation{Wyss Institute for Biologically Inspired Engineering, 60 Oxford Street, Cambridge MA 02138}

\begin{abstract}
Termite colonies construct towering, complex mounds, in a classic
example of distributed agents coordinating their activity via
interaction with a shared environment. The traditional explanation for
how this coordination occurs focuses on the idea of a ``cement
pheromone'', a chemical signal left with deposited soil that triggers
further deposition. Recent research has called this idea into
question, pointing to a more complicated behavioral response to cues
perceived with multiple senses. In this work, we explored the role of
topological cues in affecting early construction activity in
Macrotermes. We created artificial surfaces with a known range of
curvatures, coated them with nest soil, placed groups of major workers
on them, and evaluated soil displacement as a function of location at
the end of one hour. Each point on the surface has a given curvature,
inclination, and absolute height; to disambiguate these factors, we
conducted experiments with the surface in different orientations. Soil
displacement activity is consistently correlated with surface
curvature, and not with inclination nor height. Early exploration
activity is also correlated with curvature, to a lesser
degree. Topographical cues provide a long-term physical memory of
building activity in a manner that ephemeral pheromone labeling
cannot. Elucidating the roles of these and other cues for group
coordination may help provide organizing principles for swarm robotics
and other artificial systems.
\end{abstract}
\maketitle

\section{Introduction}

Termites in the Macrotermitinae subfamily construct towering earthen
structures around their nests, which provide protection and help to
ventilate and regulate internal climate \cite{Ocko2017}. The
construction of these mounds requires the translocation of hundreds of
kilograms of soil annually \cite{Turner2006}. The soil is moved from
deep beneath the colony, up through the mound, to be placed on the
outer surface of the mound \cite{Turner2006}. Soil may not be moved
directly from the subterranean source to the ultimate site of
deposition, but appears to be conveyed in stages as it travels upwards
to the surface through the action of a number of individuals
\cite{Turner2006}. The process of moving this soil leads to extensive
remodeling of the tunnels and chambers within the mound
\cite{Ocko2017}.

The mechanisms that govern where soil is removed from as well as added
to the walls of the tunnels within the mound are unknown. It has been
long thought that social insects organize their labor through a system
of indirect communication known as stigmergy \cite{Grasse59,
  Bruinsma1979, Theraulaz1999, Khuong2016}, in which the insects
manipulate a shared environment, which thereby stores information they
use to coordinate their joint activity. In the classic formulation, a
putative cement pheromone is hypothesized to be added to soil pellets
deposited by workers; other workers encountering this pheromone
respond by depositing additional soil, leading to a positive feedback
loop that results in the accumulation of soil at focal locations
\cite{Grasse59, Khuong2016}. This emphasis on cement pheromone has
been challenged by a growing number of studies in termites
\cite{Bardunias2010, Fouquet2014, Petersen2015, Green2016} and ants
\cite{Bruce2016}; it is also noteworthy that no cement pheromone has
yet been identified. Instead, other cues have been identified as
playing a role in organization of insect construction
activity. Tactile cues have been suggested to elicit deposition
responses in termites that could result in mound building
\cite{Fouquet2014}. In {\it Macrotermes}, aggregation of actively
digging workers has been shown to be a cue for recruiting additional
workers, and excavation sites shown to act as a template for
deposition \cite{Green2016}. In {\it Coptotermes formosanus}
(Shiraki), both excavation and deposition have been shown to occur at
depressions within tunnel walls depending on the behavioral state of
the termite \cite{Bardunias2009}. Bends in tunnels have also been
shown to act as cues for excavation and deposition \cite{Lee2008}.

In the language of this theme issue, the termites and the structures
they build act as a hybrid liquid/solid brain. The termites constitute
the liquid component, moving freely through the mound, interacting and
responding to local stimuli; the structures constitute the solid
component, presenting a fixed record of activity that is modified over
longer time scales. Together they form a system of distributed
cognition. The output of the actions of individual termites involved
in the construction process is not ephemeral, but remains encoded in
the topography of the structures they create. These structures thus
act as a shared memory for the termite work force. Later construction
occurs at sites that are artefacts of the labor of earlier
construction. Building activity is shaped by a combination of existing
topography, environmental conditions within the mound, and the
behavioral motivation of the workers themselves as they seek to move
and deposit wet soil or alternately excavate soil to be transported
elsewhere. The interplay of the liquid and solid aspects of the
termite collective brain provides termites with the ability to rapidly
respond to changing needs and environmental challenges, while
maintaining a structure that remains effective over the life span of
many individual termite workers.

{\it Macrotermes} may be responding to topographical cues in choosing
sites for excavation and/or deposition. Tunnel curvature may provide
cues analogous to those shown to be informative in {\it C. formosanus}
\cite{Bardunias2009, Lee2008, Lee2008b}. However, previous studies
with subterranean termites have been conducted in horizontal planar
arenas, which exclude other possible cues that may be present in freer
environments. For instance, incline (slope with respect to the
vertical) may provide a cue; similarly, termites may express geotaxis,
exhibiting a tendency to move up or down with respect to gravity to
excavate or place soil. Herein we show that early termite construction
activity is strongly associated with surface concavity (high positive
curvature, using the mean curvature definition), and displays
inconsistent correlation with inclination or geotaxis.

\section{Methods}
\subsection{Experiments}
We conducted experiments in April 2017 at the Cheetah View Field
Biology Station near Otjiwarongo, Namibia ($20^{\circ}$ $25^{\prime}$
S, $17^{\circ}$ $4^{\prime}$ E). We studied termites of the species
{\it Macrotermes michaelseni} (Sjostedt) (Blattodea: Termitidae),
placing workers on a shaped surface and recording their activity.

In order to determine the effect of the three factors noted above
(curvature, inclination and geotaxis), we designed and 3D-printed a
specialized test surface (Figure \ref{MainResults}, top row; SI Figure
\ref{ChipDesign}). The curvature varies continuously over this
surface, with areas that are concave, convex, or flat; similarly, each
point on the surface is associated with a specific inclination and
height when the surface is mounted in a given orientation (Figure
\ref{MainResults}, each column). To disambiguate the three factors, we
performed experiments with the surface mounted in three different
orientations; reorienting the surface changes the inclination and
height at each point but not the curvature. These different
orientation conditions were termed {\it horizontal}, {\it 45 degrees},
and {\it vertical} (Figure \ref{MainResults}, top row left to
right). The surface was also designed with homogeneous small
perforations, to help ensure uniform soil moisture across the surface.

For each treatment we performed experiments using 3 different
colonies, with three replicas for each colony (except for colony 2,
where {\it 45 degrees} orientation had five replicas, and {\it
  vertical} orientation had two replicas). For each experiment we
extracted termites and recently manipulated soil from an active
building site of one of the 3 colonies. After collection, termites
were stored in an enclosed container with wet paper towels, for no
longer than 6 hours before an experiment. We coated the test surface
with a layer of the recently manipulated soil approximately 2mm
thick. Mound soil was mixed with water to achieve a loose consistency
that was poured onto the surface of the manifold. This was repeated
three times to achieve the proper soil depth, and if the surface
appeared too dry during the process, water was sprayed on the surface
to ensure the soil was in a liquid Atterberg state
\cite{Atterberg}. Liquid soil on the manifolds was allowed to air-dry
for 20 minutes to allow soil to dry to its plastic Atterberg state:
approximately 25\% water by weight. Soil from manifolds coated and
dried in this manner was tested using the ASTM Standard D 4318 test to
verify that the soil had dried from its liquid Atterberg state to the
plastic state. No special effort was made to remove larger
particulates from the soil, resulting in a more granular surface. By
not sieving larger soil particles, we provided termites with soil
conditions typical of their natural environment. These particles could
potentially influence behavior, but they were distributed haphazardly
over the whole surface, with no consistent locations across trials,
and not likely to confound any effect of topology. The last row of
Figure \ref{MainResults} displays initial state examples for each
treatment.  In an indoor environment, we oriented the test surface
according to the experimental condition, placed 25 termites of the
major worker caste atop it, and enclosed it in a custom-built acrylic
box, illuminated by LED lamps on two sides and with a wet cloth
underneath the surface to maintain air humidity. Experiments were
filmed (1080x1920 resolution, 30 frames per second) from above for an
hour, or, given that there were no physical boundaries to the surface
and termites occasionally fell off its edges, until no termites
remained on the surface.

\subsection{Analysis}
Since the surface was designed digitally, we could easily calculate
the curvature, height, and inclination of every point for each
orientation. A geometric transformation gave the correspondence
between each pixel in the video image and the corresponding point on
the surface. We used the mean curvature definition, $H=(k_1+ k_2)/2$,
where $k_1$ and $k_2$ are the two principal curvatures of the surface
at a point. A principal curvature is positive if it curves in the
direction of the surface normal vector and negative if opposite to
it. Inclination was defined as the modulus of the gradient of the
tangent plane at a point.

To bring all videos into consistent registration, we performed the
following procedure. First, we averaged all frames over each minute of
the video (30 frames per second, or 1800 frames per minute), providing
the equivalent of a long exposure photo for each minute of the
video. This provided 60 averaged images (for experiments that did last
the full hour). Termites that moved during each one-minute period were
averaged away and were in most cases invisible in these averaged
images (SI Figure \ref{LongExposure}).

Next, we used an automatic image stabilization algorithm provided by
Matlab (Matlab2016a, Video Stabilization Using Point Feature Matching
package) to eliminate camera drift that occurred over the course of
some experiments. Removing termites before this step was necessary to
avoid false automatic features from being selected and affecting the
image stabilization. Next, to adjust for small inconsistencies in
camera positioning between videos (orientation, position, and height
relative to the surface), we manually annotated the edge of the
surface in each video to define a region of interest (ROI), and
applied an algorithm to match ROIs for all trials with the same
treatment: resize to have the same overall area, shift to have the
same centroid, and finally test possible rotations in order to
maximize the overlap between them.

To quantify soil displacement, for each replica we averaged the first
and last 5 minutes of video into single images (respectively first and
second column of SI Figures
\ref{Replicas_Horizontal}--\ref{Replicas_Vertical}) and performed a
background subtraction by taking the difference between these initial
and end states. This procedure removed visible features in the soil
covering the surface (e.g., small stones) that remain unchanged since
the start of the trial. The remaining image provided a record of soil
displacement activity, including both excavations and
depositions. Finally, this RGB image was converted to monochrome
(third column of SI Figures
\ref{Replicas_Horizontal}--\ref{Replicas_Vertical}), and the results
of the multiple replicas all averaged into a single image, with pixel
brightness representing the strength of consistency of soil
displacement at each point. Averaging over the multiple trials for
each orientation removed any remaining termites that did remain
immobile (and therefore visible) in the last minutes within a single
replica (Figure \ref{MainResults}, row 6).

The same factors of curvature, inclination, and height may also affect
where termites prefer to move in addition to where they prefer to dig,
and the former effect may contribute to the latter. To investigate how
these factors influence termite movement, we manually marked termite
locations in each video at every 10 seconds for the first 5 minutes of
each recording (for a total of 40 frames per replica). This time range
was chosen in order to capture initial exploration activity, during a
period before excavations began, since active excavation influences
activity of other termites \cite{Green2016}. Since termites sometimes
do not start moving for some time after being placed on the surface,
we excluded termites that had not yet moved to a distance of at least
30 pixels from their positions in the first tracked frame. One replica
of the {\it 45 degrees} orientation was not tracked due to the first
minutes of the video of that replica being lost. Each marked termite
position was translated to a circle of approximately the size of a
termite (15 pixels radius), and all circles superimposed for the
tracked period of a given replica, to produce a heatmap indicating
termite locations during early exploration of the surface (Figure
\ref{MainResults}, row 5; SI Figures
\ref{Replicas_Horizontal}--\ref{Replicas_Vertical}, last column).

The averaged initial state, averaged end state, soil displacement
(construction activity), and termite early exploration (movement
activity) for each replica of all orientations can be seen
respectively through the 4 columns of SI Figures
\ref{Replicas_Horizontal}--\ref{Replicas_Vertical}.

\section{Results}

\begin{figure}
  \centering
  \includegraphics[width=0.6\textwidth]{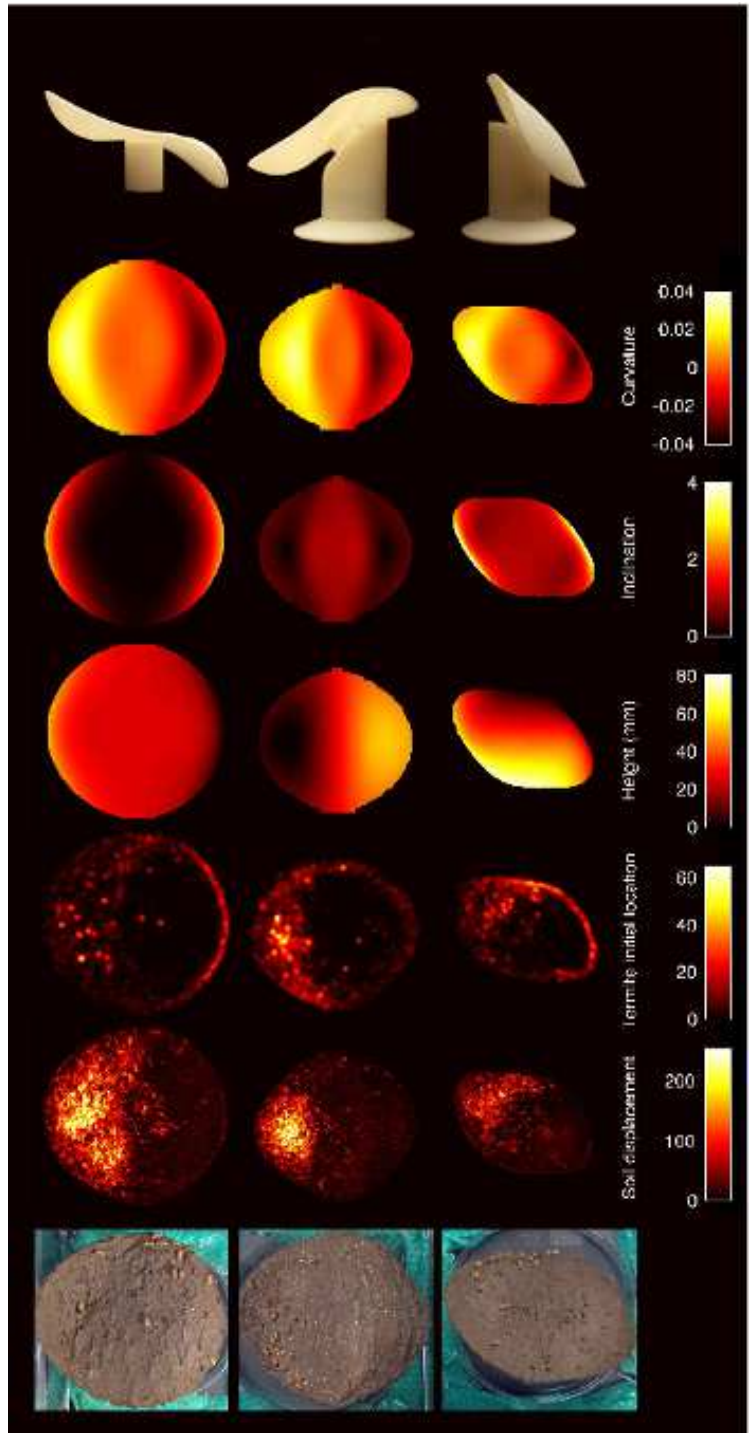}
  \caption{3D-printed specialized surface (top row), with regions that
    are concave, convex or flat. Each column represents a different
    reorientation of that same surface, and were termed as: {\it
      horizontal}, {\it 45 degrees}, and {\it vertical}. Rows 2 to 4
    display the heatmap of the three analyzed variables (curvature,
    inclination and height, respectively). The 5th row shows the
    heatmap of early termite exploration, and the 6th row shows the
    averaged soil displacement activity; a consistent correlation of
    activity with curvature, and not with inclination or height, is
    visible. Lastly, the bottom row shows a snapshot of the initial
    state of one experiment per orientation.
    \label{MainResults}}
\end{figure}

\begin{figure}
  \centering
  \includegraphics[width=0.95\textwidth]{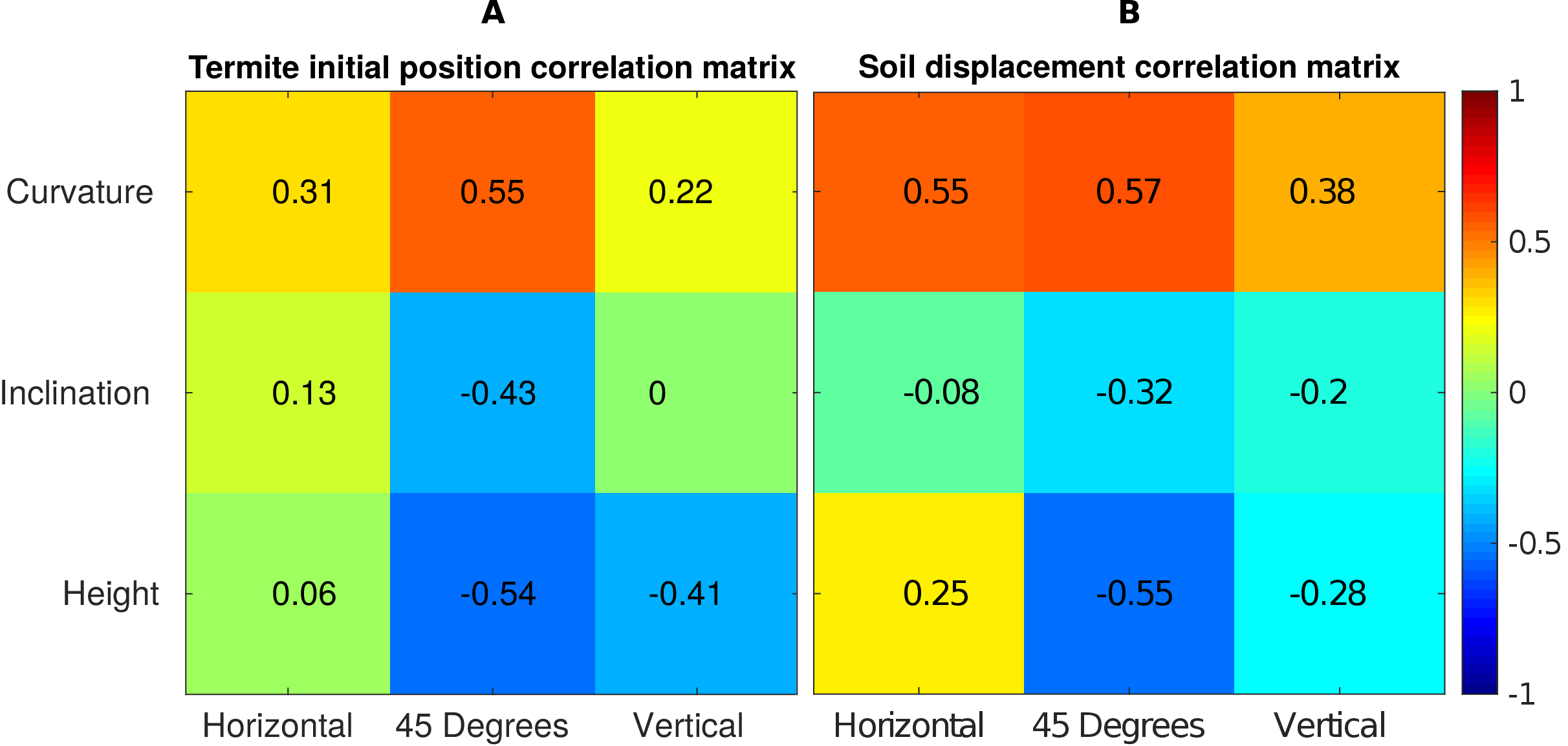}
    \caption{Correlation matrices of curvature, inclination and
      height, in relation to: {\bf A}, termite locations during early
      exploration; {\bf B}, soil displacement activity. Both matrices
      show that curvature is the consistent and sole driver, among
      these geometric factors, of these two forms of
      activity. \label{Correlation}}
\end{figure}

Figure \ref{MainResults} (second to sixth row) shows the different
results discussed above as a function of position on the surface,
displayed as a heatmap. Columns represent the three different
orientations (as seen in the first row), and the following rows
represent curvature, inclination, height, termite early movement, and
construction activity. The correlation of curvature with construction
activity, and the lack of consistent correlation of inclination or
height with construction activity, is visible across all
orientations. Density plots and averages of activity as a function of
curvature for all orientations are shown in SI Figure
\ref{Density}. While excavation and deposition cannot be separately
quantified with our method, visual inspection showed that deposition
occurred primarily in the immediate vicinity of excavation sites (SI
Figures \ref{Replicas_Horizontal}--\ref{Replicas_Vertical}),
consistent with previous studies \cite{Green2016}. Figure
\ref{Correlation} quantifies the correlation matrices between surface
geometric factors (curvature, inclination and height) and both termite
early locations and construction activity.

These results show that curvature is the consistent and sole driver,
among the measured geometric candidates, of both early termite
positioning and construction activity. Curvature is more strongly
correlated with soil displacement than with termite locations,
indicating that the observed preference for construction in concave
areas is not solely due to (though possibly influenced by) the
preference for spending time in those places.

It is worth noting that these correlation values are at best a lower
bound estimate of the agreement between activity and curvature. The
curvature has a continuous distribution of values, while termite
activity happens in relatively small regions of the surface, with much
of the surface displaying no activity.

\section{Conclusions}

Construction in termites requires that each pellet of soil used in
building must first be excavated from someplace within the nest or
beyond \cite{Bardunias2009}. {\it Macrotermes} mounds are permeated
with a system of conduit tunnels that allow passage as well as
affecting airflow to the nest and fungus garden below. Our study
suggests that the curvature of these tunnels within the mound acts as
a cue that organizes their construction, in a physical feedback
loop. Termites, the mobile agents that act as the ``liquid'' component
in the liquid/solid brain analogy, behave like a liquid in a further
sense, pooling in concave areas. Construction activity that then
occurs starting in these zones further shapes the solid component
embodied by the mound. These ongoing changes affect the local
perceptions the termites have, and continue to influence the possible
outcomes of the growing structure.

Zones of tunnel with high rates of curvature can occur as artifacts of
tunnel excavation, and may be similar to the depressions in tunnel
walls that have been shown to attract labor in {\it C. formosanus}
\cite{Lee2008}. The manner in which termites assess curvature in
tunnel walls is unknown, but presumably involves a combination of
antennation and proprioception of body posture \cite{Staudacher2005,
  Bardunias2009b}. Lee et al. \cite{Lee2008} found that termites
excavated at depressions in tunnel walls when their forward progress
was impeded.

The rate of curvature of tunnels can be expected to be low for the
large, smooth tunnels that make up the main air-flow conduits in the
central portion of the mound \cite{Turner2008}. Where tunnels abruptly
turn back near the outer surface of the mound, or in the narrowing at
the top of the mound, high curvature may focus excavation, expanding
the internal conduits, or soil deposition, sealing conduits and
thickening the mound walls.

An attraction to curvatures with values typical of excavation sites
may lead to a positive feedback loop of activity, eventually producing
a tunnel originating at such a site. Alternatively, when encountering
high curvatures near transitions in the topography, termites could
react by smoothing away these features through either excavation or
deposition (SI Figure \ref{Transitions}). These two possibilities
represent computations the liquid/solid brain can perform, amplifying
or smoothing out initial irregularities in tunnel walls. In future
work, we will develop simulation models to further explore
quantitatively the ways in which these different types of computations
occur under different conditions, and how they might contribute to
mound function. The existence of a dynamic equilibrium between
opposing excavation and deposition behavior at the same sites has been
shown in {\it C. formosanus} to alternately expand or cover over
depressions in tunnel walls \cite{Bardunias2009, Lee2008}. The same
cue elicits opposing digging and filling based on the prior behavioral
state of the termites, loaded with soil vs seeking a digging
site. This system allows for a rapid response by the termites to
changing conditions that alter the behavioral state of the termite
workers. A local imbalance in workers seeking to excavate soil or
loaded workers looking to deposit will govern what occurs at zones of
high curvature.

This study is part of an ongoing effort to elucidate the cues termites
use in collective construction, identifying the factors in their
environment and their interactions with others that shape their
behavior. Ascertaining these behavioral rules, and relating them to
characteristics of the mounds the insects build together, will help
advance our understanding of how individual agent actions connect to
collective colony outcomes. Connecting these small-scale and
large-scale levels of description has the potential to help us
understand the operation of other social insect species, as well as
providing principles for the design of artificial swarm systems, where
large numbers of independent robots follow simple rules in order to
build structures or achieve other desired outcomes
\cite{Brambilla2013}.

\section{Author Contributions}

Study concept: ST, RN, JW; Study design: PB, DSC, NC, ST, RN, JW;
field/laboratory work: PB, NC, DSC; data analysis: DSC;
Interpretation: DSC, PB, NC, ST, RN, JW; writing: DSC, PB, JW; All
authors gave final approval for publication.

\section{Acknowledgments}

Research reported in this publication was supported by the National
Institute of General Medical Sciences of the National Institutes of
Health under award number R01GM112633. The content is solely the
responsibility of the authors and does not necessarily represent the
official views of the National Institutes of Health. We thank the
Cheetah Conservation Fund for providing research facilities.

\newpage

\appendix
\section{Supplementary Information}
\setcounter{figure}{0}    
\counterwithin{figure}{section}

\begin{figure}[hbt]
  \centering
  \includegraphics[width=0.7\textwidth]{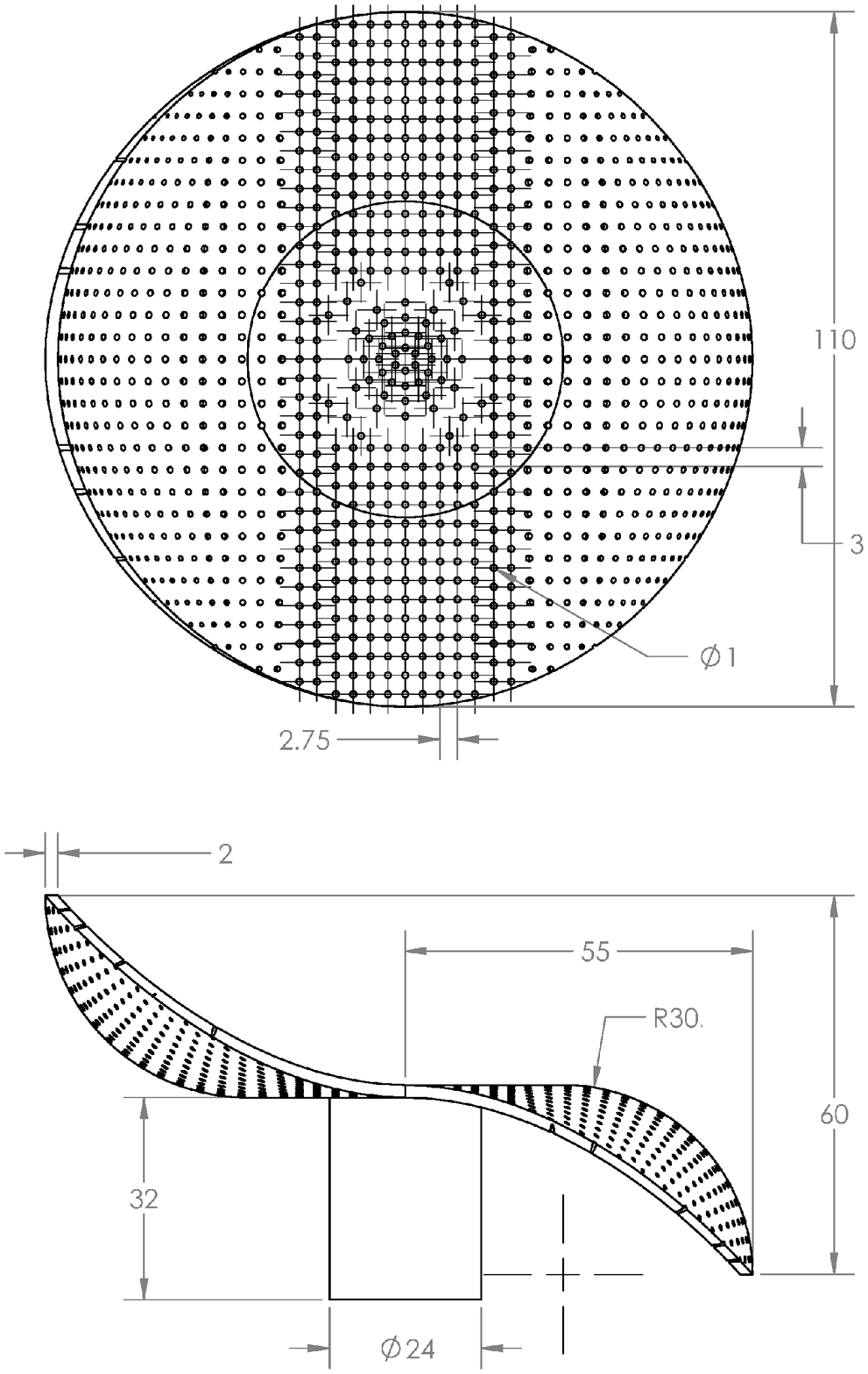}
  \caption{Design detail for the test surface used in the
    experiments. All dimensions in millimeters. \label{ChipDesign}}
\end{figure}

\begin{figure}
 \centering
  \includegraphics[width=0.7\textwidth]{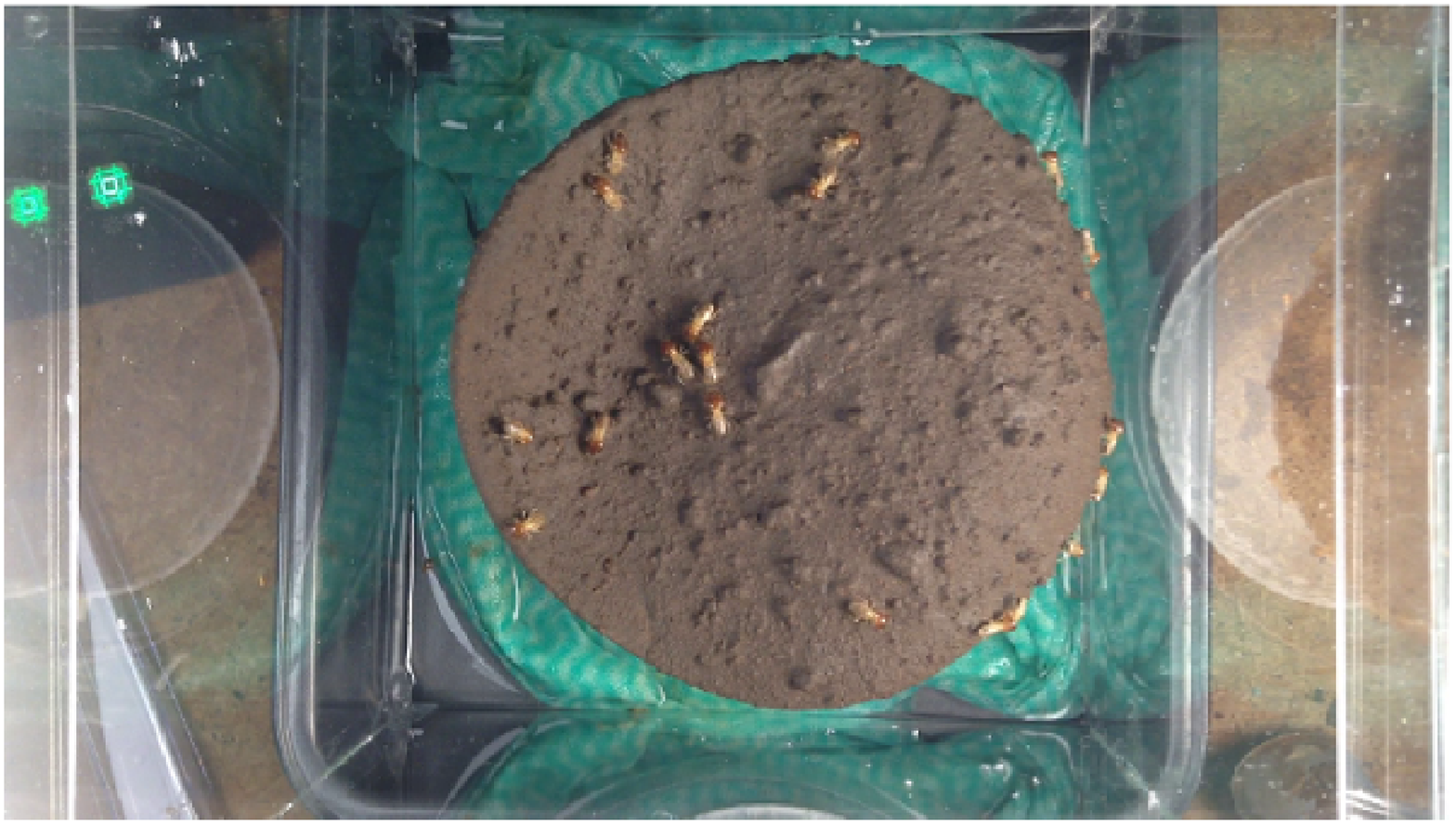}\\
  \includegraphics[width=0.7\textwidth]{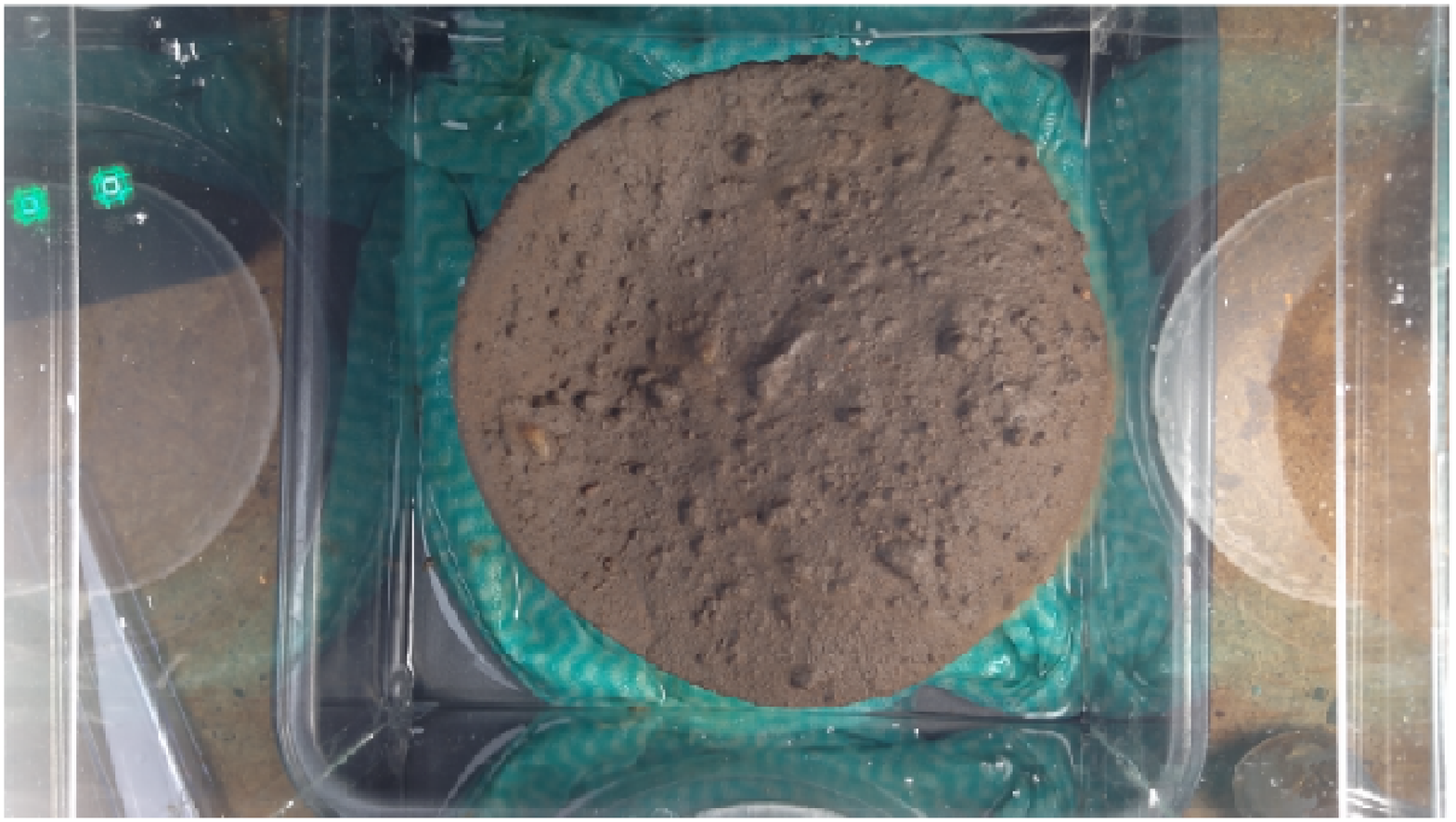}\\
  \caption{Example of the long exposure method used to minimize
    termite presence in the frames analyzed: top photo was taken at
    5 min and 30s of one of the experiments; bottom photo was taken
    averaging the 1800 frames from minute 5 to minute 6 of the
    video. \label{LongExposure}}
\end{figure}

\begin{figure}
  \centering
  \includegraphics[height=0.8\textheight]{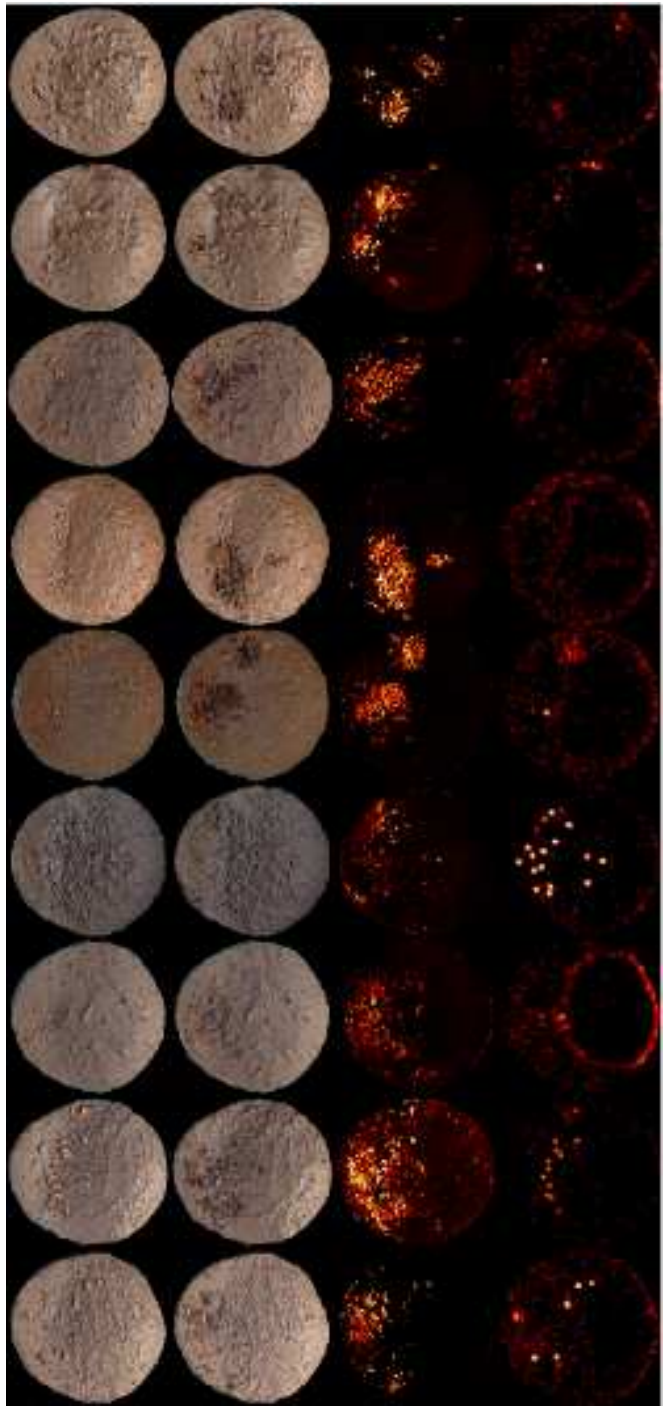}
  \caption{{\it Horizontal} orientation, all replicas. Each row represents a
    different experiment, and columns are respectively: initial state
    (averaged image over 5 minutes), end state (averaged image over 5
    minutes), construction activity (obtained by subtracting column 1
    from column 2) and termite locations during the initial 5 minutes of the
    experiment. \label{Replicas_Horizontal}}
\end{figure}

\begin{figure}
  \centering
  \includegraphics[height=0.8\textheight]{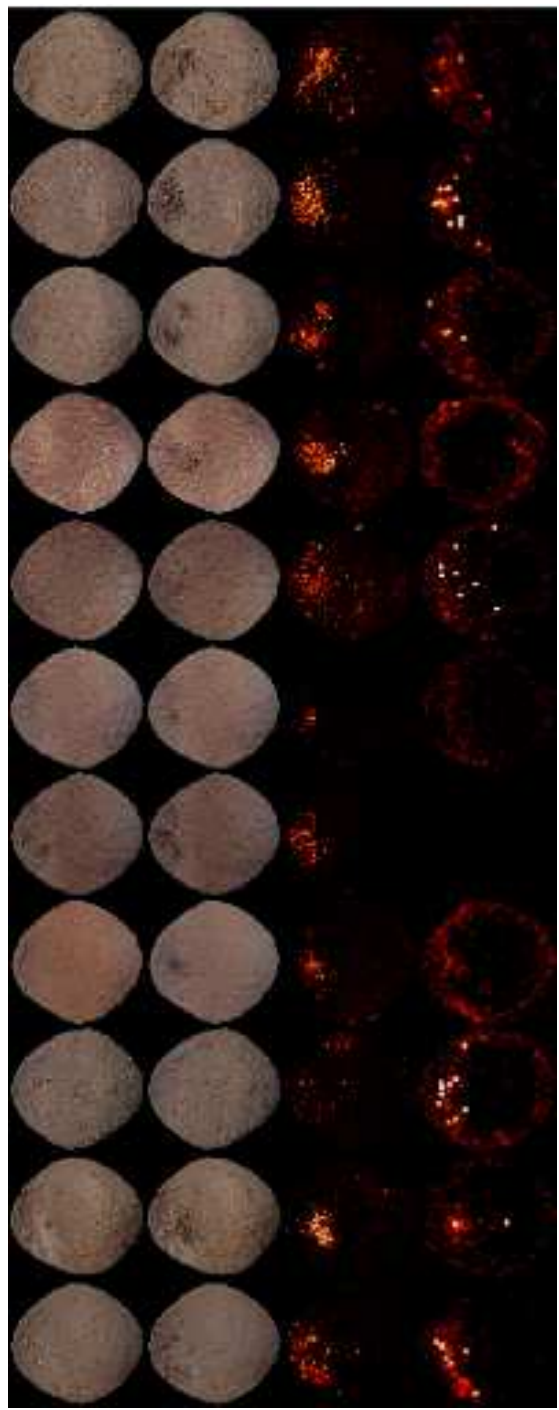}
  \caption{{\it 45 Degrees} orientation, all replicas. Each row
    represents a different experiment; columns are as in Figure
    \ref{Replicas_Horizontal}. \label{Replicas_45}}
\end{figure}

\begin{figure}
  \centering
  \includegraphics[height=0.8\textheight]{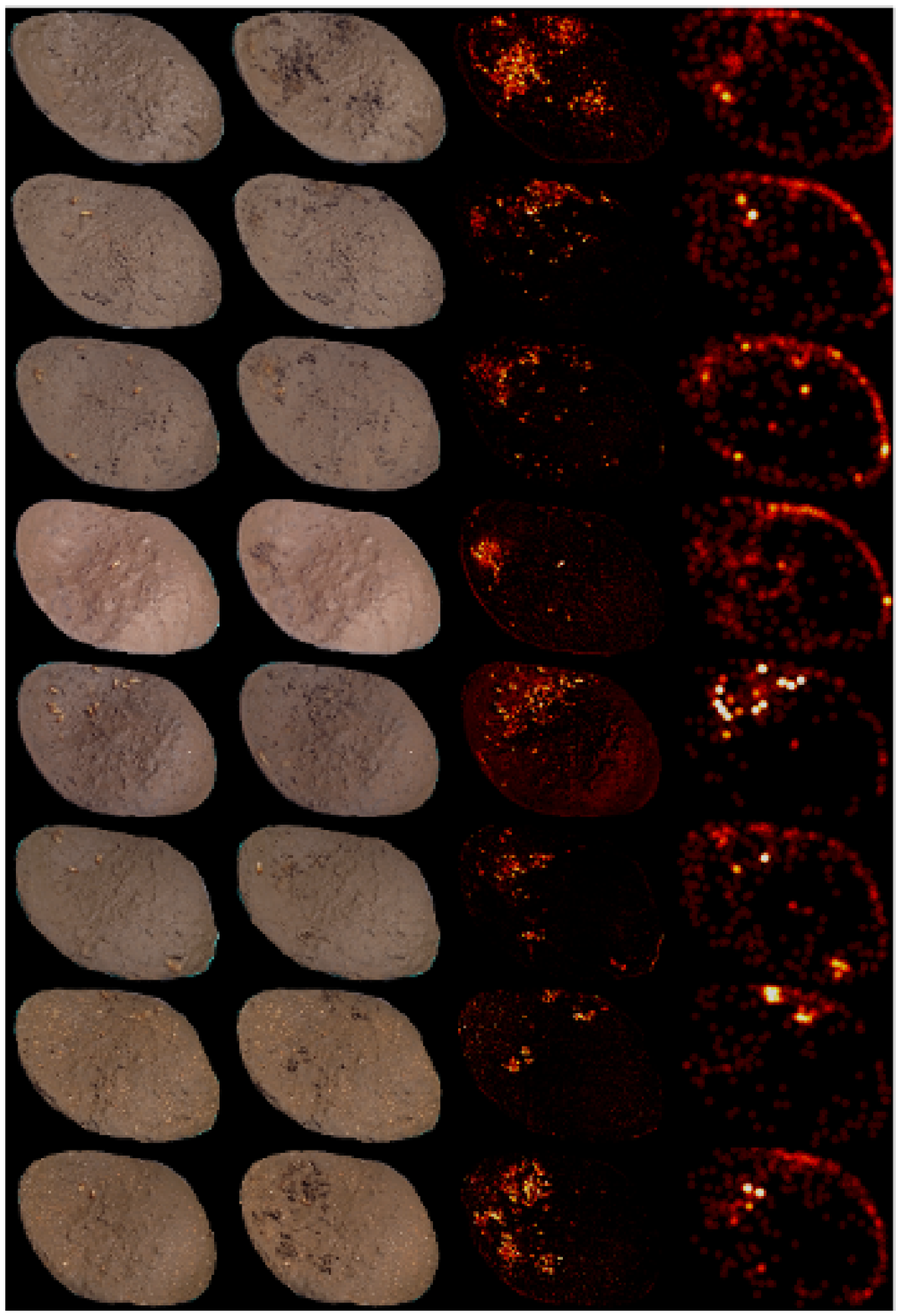}
  \caption{{\it Vertical} orientation, all replicas. Each row represents a
    different experiment; columns are as in Figure \ref{Replicas_Horizontal}. \label{Replicas_Vertical}}
\end{figure}

\begin{figure}
  \centering
  \includegraphics[width=0.48\textwidth]{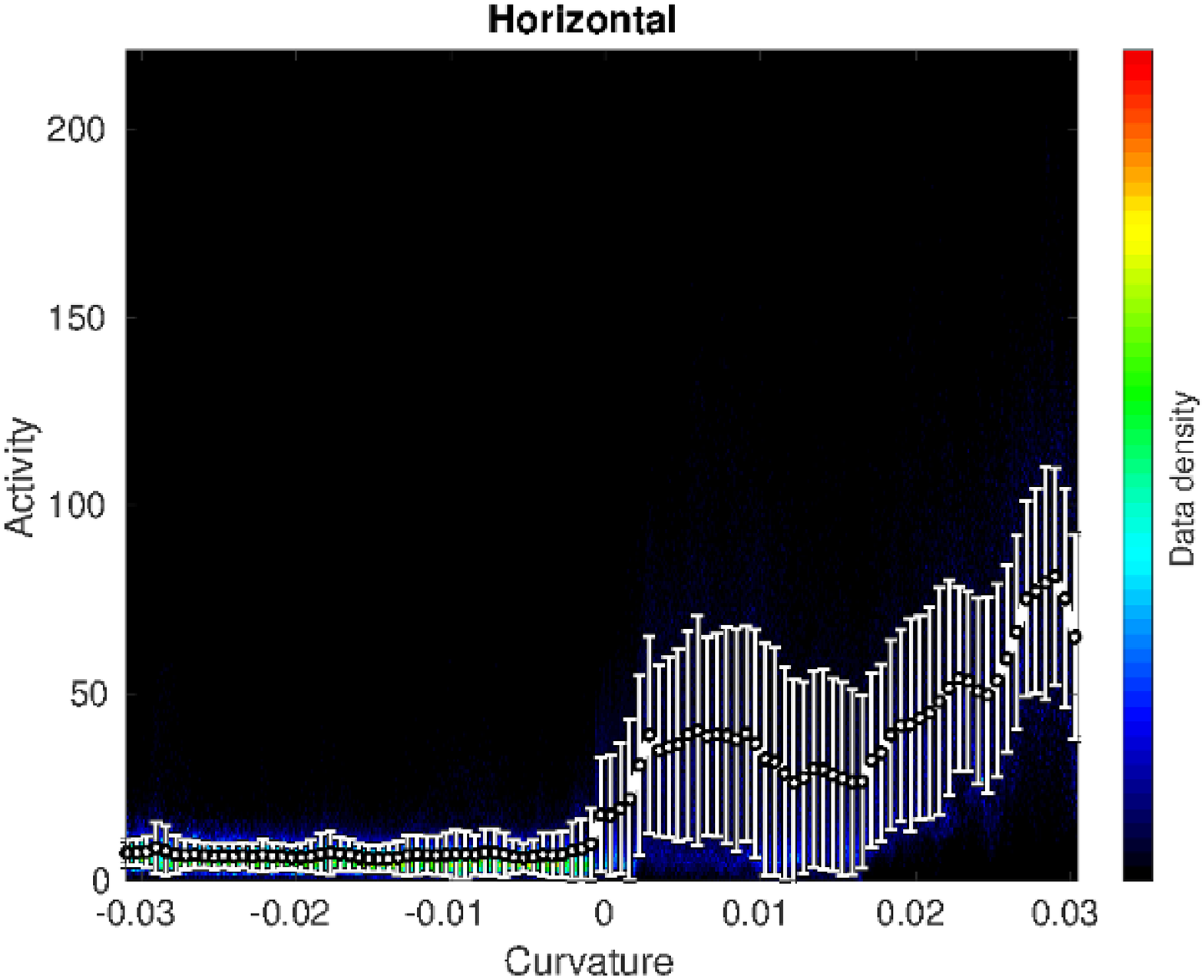}
  \includegraphics[width=0.48\textwidth]{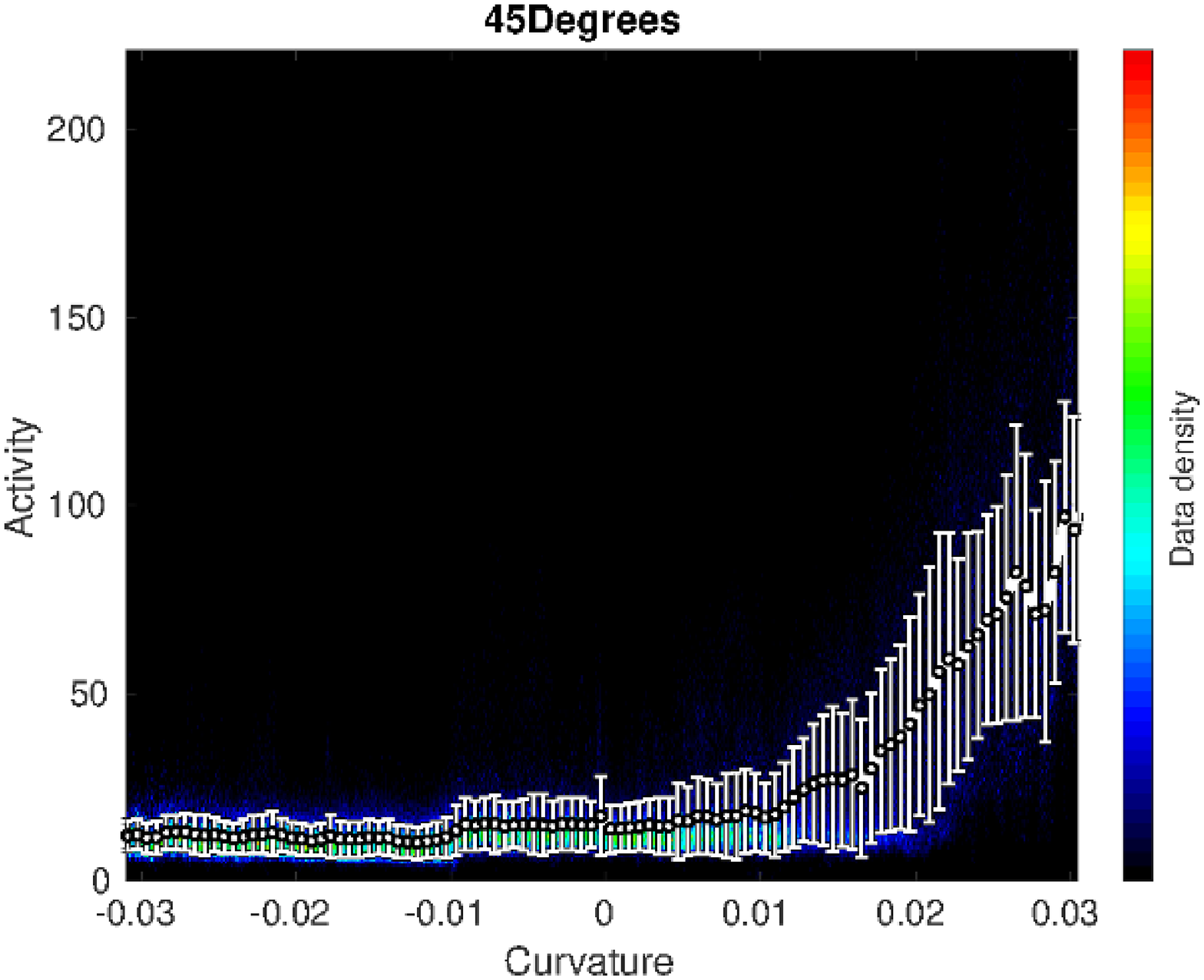}\\
  \includegraphics[width=0.48\textwidth]{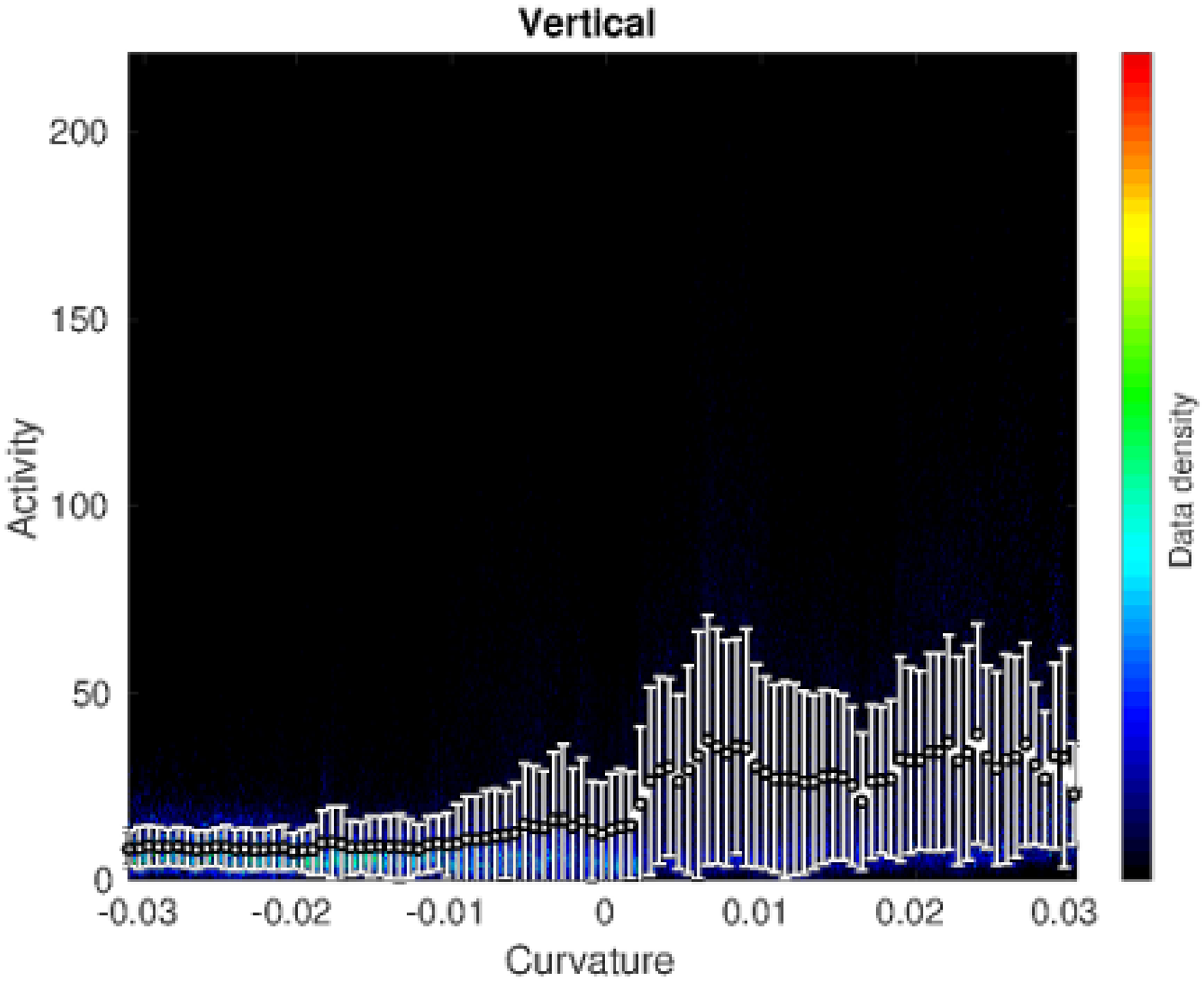}
  \caption{Termite activity as a function of curvature for each of the
    orientations. White lines and bars represent average activity and
    their corresponding standard deviations. Colors display the data
    density (normalized by data quantity for each
    curvature.) \label{Density}}
\end{figure}

\begin{figure}
  \centering
  \includegraphics[width=0.7\textwidth]{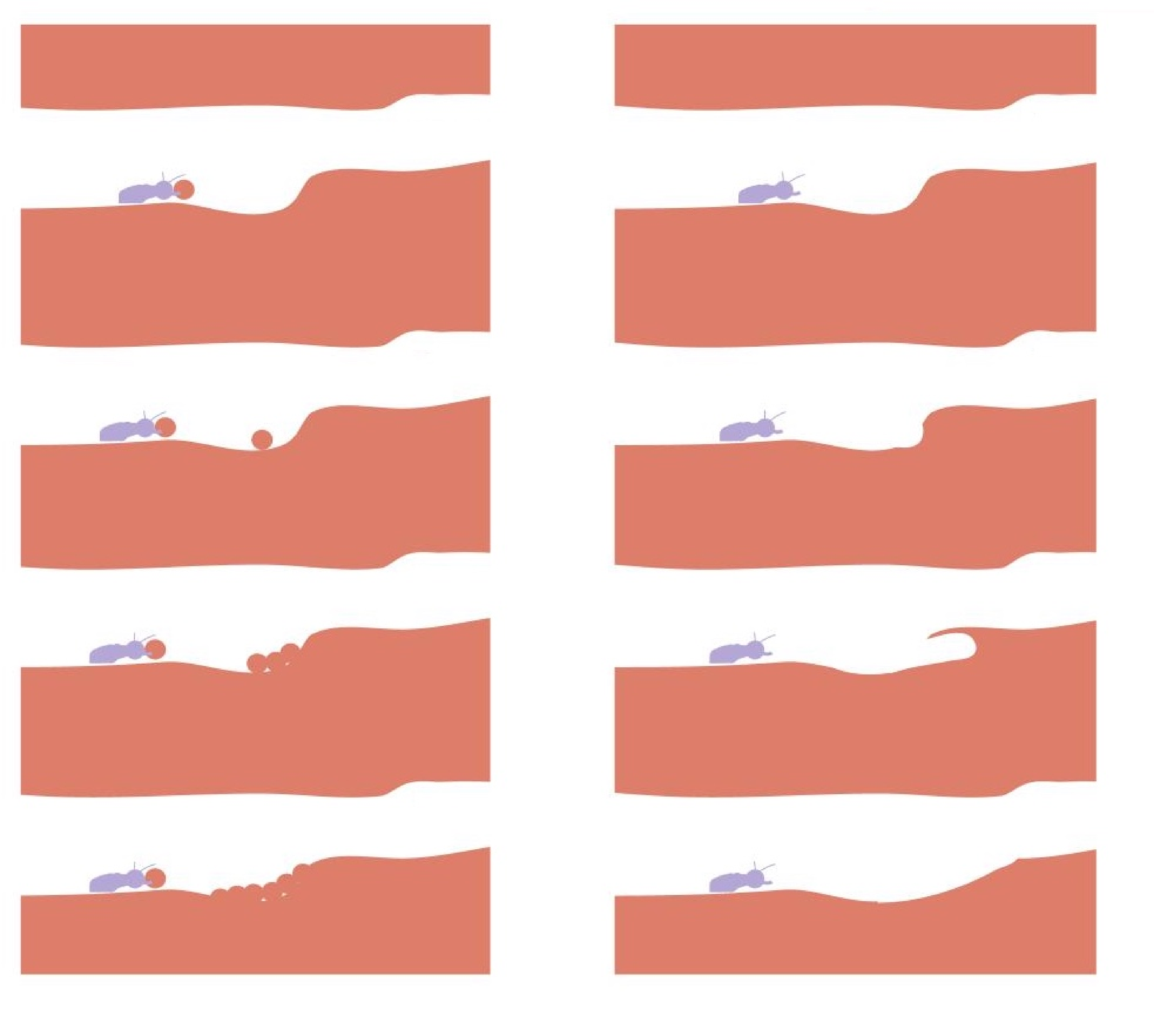}
  \caption{Example of how activity at high-curvature zones can lead
    to smoothing out of features. \label{Transitions}}
\end{figure}

\end{document}